\documentclass[10pt]{iopart}
\usepackage{epsf}
\usepackage{iopams}
\usepackage{subfigure}
\begin{document}

%
\title{Leading off-diagonal contribution to the
       spectral form factor of chaotic quantum systems}

\author{Marko Turek and Klaus Richter}

\address{Institut f{\"u}r Theoretische Physik,
           Universit{\"a}t Regensburg,
           D-93040, Germany}

\ead{marko.turek@physik.uni-regensburg.de}

\date{\today}

\begin{abstract}
We semiclassically derive the leading off-diagonal correction 
to the spectral form factor of quantum systems with a chaotic 
classical counterpart. To this end we present a phase space 
generalization of a recent approach 
for uniformly hyperbolic systems \cite{Sieber01,Sieber02}.
Our results coincide with corresponding random matrix
predictions. Furthermore, we study the transition from the Gaussian 
orthogonal to the Gaussian unitary ensemble.
\end{abstract}
\vspace*{-5mm}




%

Advanced semiclassical methods have been very successful
to significantly improve our understanding of complex, classically
chaotic quantum systems\cite{BooksQC,Richter00}. This holds particularly
true for observables which can be deduced from the (single-particle) 
Green function of the quantum system, such as the density of states, 
photo absorption, or orbital magnetism, to name a few.
However, for quantities which are based on Green function products 
the situation is much more involved. These include linear
response functions, e.g.\ for quantum transport, spectral correlation 
functions or, more generally, spectral statistics. A semiclassical
treatment of such quantities is usually faced with the serious problem
of evaluating multiple infinite sums over phase-carrying classical paths,
which arise from the semiclassical representation of the Green functions 
in the limit $\hbar \to 0$. A prominent example is the
spectral two-point correlator or its Fourier transform
the spectral form factor $K(\tau)$.
Random matrix assumptions leading to the prediction of a universal form 
for $K(\tau)$ for classically chaotic quantum systems
are supported by experimental and numerical data for a vast 
number of systems from different disciplines in physics
\cite{BooksQC}.
Nevertheless, a corresponding analytical approach proving the 
conjectured \cite{Bohigas84}
universality by explicitly including
the underlying chaotic classical dynamics is still lacking.
In this respect semiclassical techniques, bridging classical
and quantum dynamics, appear to be natural tools but have to
cope with the above mentioned problems when evaluating 
spectral correlations.  For the form factor this 
involves the computation of (energy) averaged
double sums over periodic orbits which has been addressed in
several semiclassical approaches \cite{semicl}.

Recently, $K(\tau)$ was semiclassically investigated
for uniformly hyperbolic two-dimensional 
systems where the classical dynamics is governed by a single 
Lyapunov exponent \cite{Sieber01,Sieber02}.
By going beyond the usual diagonal approximation \cite{Berry85},
($K^{(1)}(\tau) \approx 2 \tau$, describing the limit
of spectral long-range correlations),
the next to leading order contribution to $K(\tau)$ was computed
for systems with time-reversal symmetry.
This was achieved by identifying 
off-diagonal pairs of correlated periodic orbits which are associated with
each other via selfcrossings in configuration space.  
Based on this orbit class the random matrix theory (RMT)
prediction for the form factor 
in the Gaussian Orthogonal Ensemble (GOE), 
$K(\tau) \approx 2 \tau - 2\tau^2$
 for $\tau \to 0$, could be derived 
\cite{Sieber01,Sieber02}, see also \cite{others}.
However, the important question remains whether
the above result, as well as the RMT prediction, are 
specific for systems with uniformly hyperbolic dynamics, or 
whether they pertain for the much broader class of chaotic
systems with different periodic orbits having different
Lyapunov exponents.

Here we present a generalization of the
semiclassical approach outlined above to 
such non-uniformly hyperbolic
systems in two dimensions and show that 
under rather general conditions the term
$\sim -2\tau^2$ in $K(\tau)$ is indeed retained.
To this end
we develop a canonically invariant approach
which is based on phase space arguments
only. We identify as the relevant 
objects 
'crossing regions' in phase space which can 
involve more than one selfcrossing in configuration
space.  We express the action differences
of the considered orbit pairs
in terms of local phase space properties, 
the directions of the stable and unstable
manifolds, and
present a method for counting
the 'crossing regions'. This allows us eventually to
determine their contribution to $K(\tau)$.

The semiclassical limit implies
a large energy $E$ compared to
the mean level spacing $1/\bar{d}(E)$.
The energy dependence of $K(\tau)$
is smoothed out by an average 
over a classically small
but quantum mechanically large
energy window of size $\Delta E$.
In the considered limit $\Delta E$ can be
chosen such that
$1/\bar{d}(E) \ll \Delta E \ll E$.
In this energy regime all classical
actions, such as the action $S_{\rm spo}$
of the shortest periodic orbit, are much larger
than $\hbar$, i.e. $S_{\rm spo}/\hbar \gg 1$.
Then one can employ Gutzwiller's trace 
formula \cite{BooksQC} for the oscillating part
of the density of 
states and evaluate the Fourier transform of the
spectral two-point correlation function.
This gives for the form factor
a double sum over periodic orbits \cite{Berry85}, 
\begin{equation}
  \label{K_tau}
  K(\tau) = \frac{1}{T_H} \sum\limits_{\gamma, \bar{\gamma}}
  \left\langle A_\gamma A^*_{\bar{\gamma}} 
  \exp\left(i \frac{S_{\gamma, \bar{\gamma}}}{\hbar}\right)
  \delta \left( T - \frac{T_{\gamma} + T_{\bar{\gamma}}}{2}
    \right) \right\rangle_{\Delta E} \; ,
\end{equation}
with the time scaled according to $\tau = T/T_{\rm H}$ where
$T_{\rm H} = 2 \pi \hbar \bar{d}(E)$ is the Heisenberg time.
We represent each periodic orbit $\gamma$ in terms of 
its phase space coordinates 
${\bi x}=({\bi q},{\bi p})$ where  ${\bi q}$ and  ${\bi p}$
are the position and momentum coordinates
in two dimensions. In (\ref{K_tau}), 
$T_\gamma$ is the period of an orbit  ${\bi x}_\gamma$
and $A_\gamma$ includes both,  its weight and Maslov index.
The action difference between two orbits ${\bi x}_\gamma$ and
${\bi x}_{\bar{\gamma}}$ is given by
$S_{\gamma,\bar{\gamma}}$.  The argument of the double sum
is a rapidly oscillating function and hence
the energy average suppresses most of the terms.
To obtain a nonvanishing contribution to $K(\tau)$ 
the action difference must be small, i.e. 
$S_{\gamma,\bar{\gamma}} \lesssim \hbar$.
Thus, for the time reversal case, the largest contribution 
is due to pairs of
one path ${\bi x}_\gamma$ with itself
or its time-reversed partner
${\bi x}_\gamma^i = ({\bi q}_\gamma(T_\gamma - t),
                    -{\bi p}_\gamma(T_\gamma - t))$
with vanishing action difference.
Including only this type of pairs, known as the diagonal
approximation \cite{Berry85}, reproduces the linear
contribution 
$K^{(1)}(\tau)\approx 2 \tau$.

\begin{figure}
\begin{center}
\hspace{2cm}
\subfigure[]{
 \label{fig_partner}
 \epsfxsize=0.38\textwidth
 \epsfbox{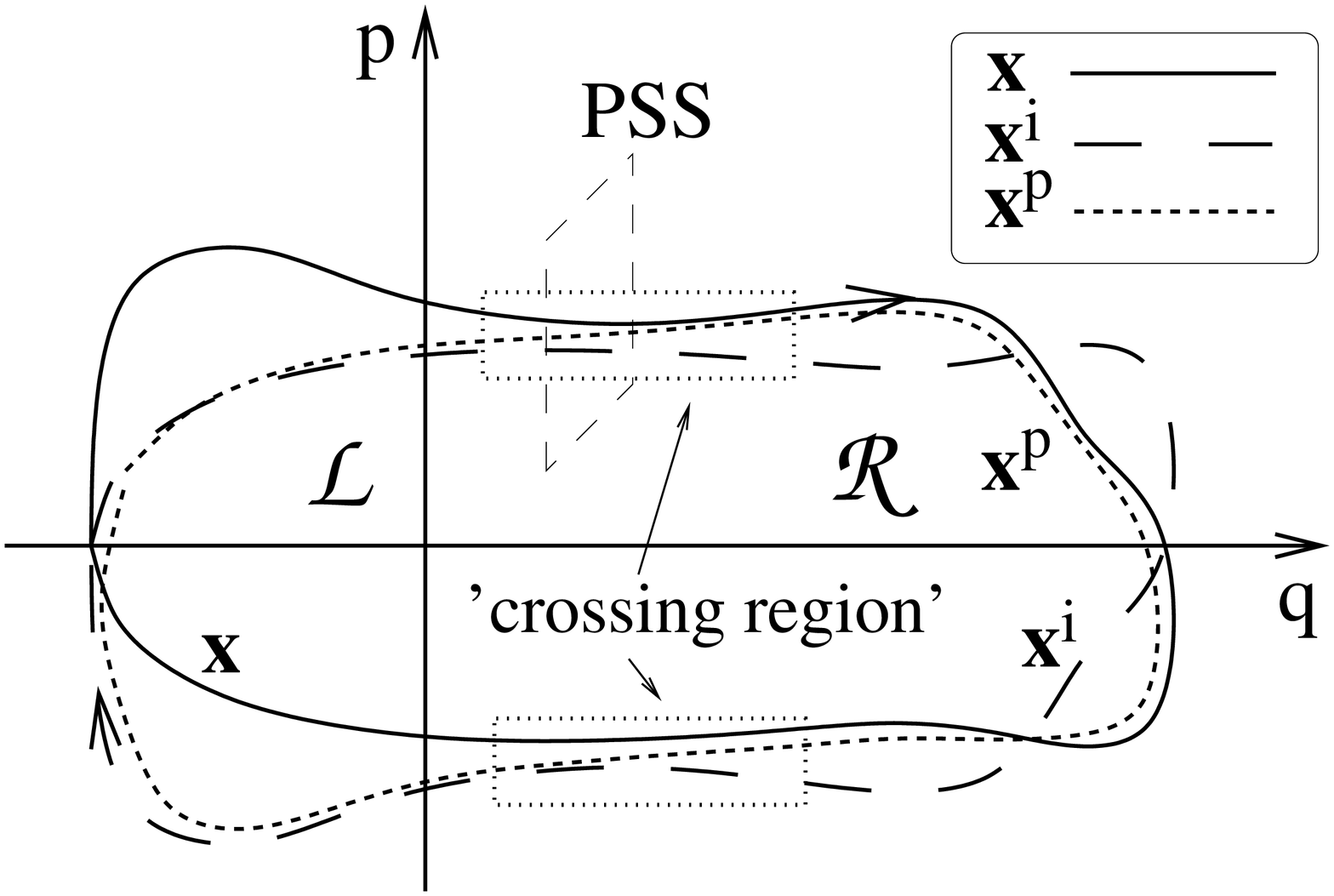}}
\hspace{3mm}
\subfigure[]{
 \label{fig_pss}
 \epsfxsize=0.38\textwidth
 \epsfbox{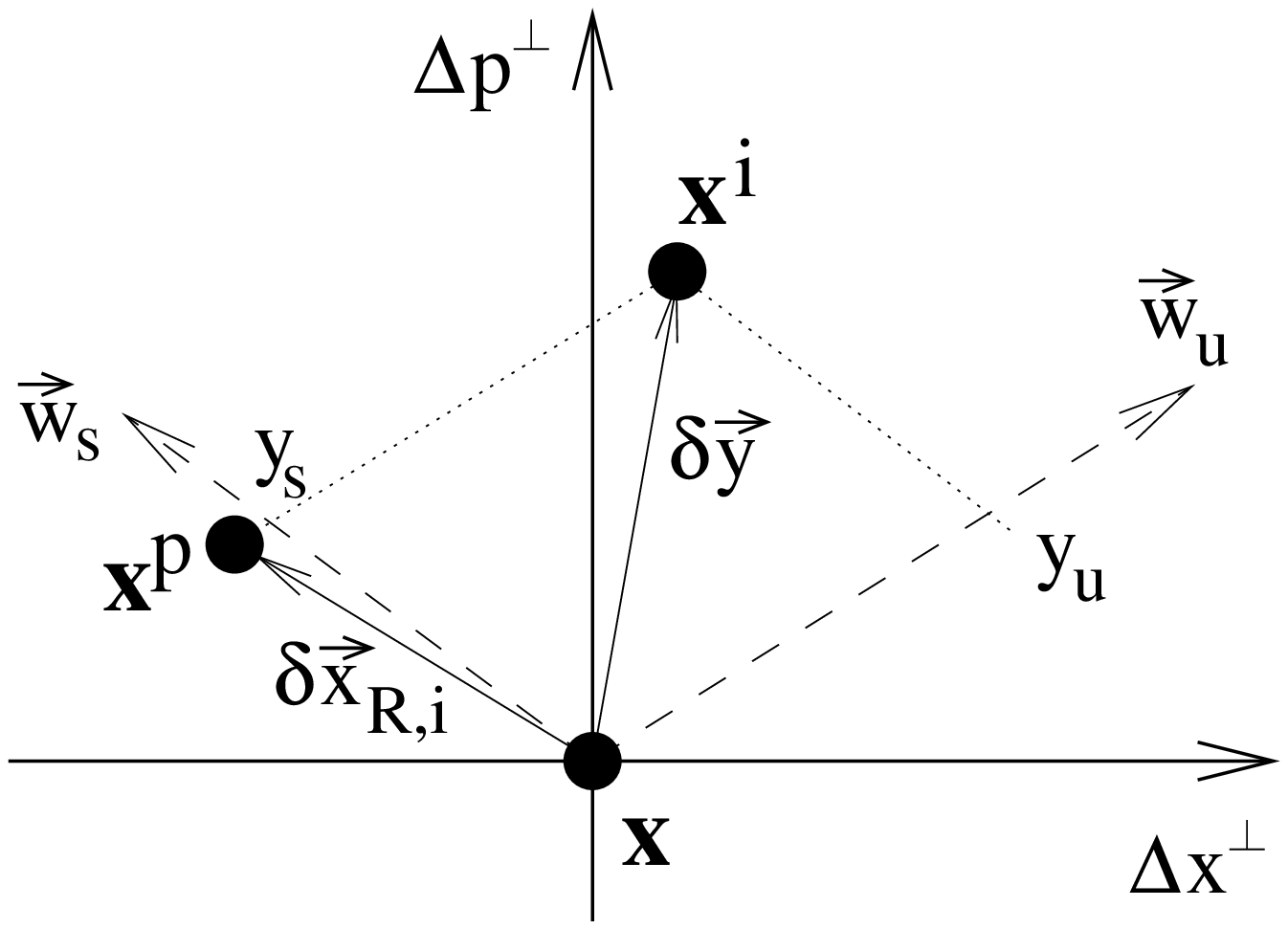}}
\end{center}
\caption{
(a)
Sketch of a correlated orbit pair in phase space
 (shown is a pro\-jec\-tion of the four-dimensional space).
 The original periodic orbit ${\bi x}$, the time reversed 
 orbit ${\bi x}^i$, and the partner orbit ${\bi x}^p$ are
 represented by the solid, dashed and dotted line, respectively.
 Due to time reversal symmetry each 'crossing region'
 appears twice. The Poincar{\'e} surface of section (PSS) defined
 by the perpendicular coordinates
 $(\Delta x^\perp, \Delta p^\perp)$ is indicated in the upper
 'crossing region'.
(b)
 PSS with\-in the 'crossing region' at the beginning
 of loop ${\cal R}$.  The path ${\bi x}$
 defines the center of the coordinate system while the vector
 $\delta \vec{y}$ points towards the time-reversed
 orbit ${\bi x}^i$. The position of the 
 partner orbit ${\bi x}^p$ is given by the vector
 $\delta \vec{x}_{R,i}$.
}
\label{figure1}
\end{figure}

In the recent approach beyond the diagonal approximation
pairs of closely related periodic orbits with a small
action difference have been constructed by analyzing
self-crossings of the orbits in configuration space
\cite{Sieber01,Sieber02}. 
Here we proceed in a different way 
by generalizing the approach via self-crossings 
to 'crossing regions' in phase space:
We will show that if a periodic orbit ${\bi x}_\gamma$
comes close to its time-reversed version ${\bi x}^i_\gamma$
in such a 'crossing region' this can imply the existence
of yet another periodic orbit
${\bi x}_{\bar{\gamma}}={\bi x}^p_\gamma$
with a small action difference $S_{\gamma,{\bar{\gamma}}}$
between the two.
The partner orbit ${\bi x}^p_\gamma$ 
follows the original ${\bi x}_\gamma$ 
in a first segment ${\cal R}$
and then the time-reversed path
${\bi x}^i_\gamma$ during the rest of the
time in the second segment ${\cal L}$,
see figure \ref{fig_partner}. 
It will turn out
that not every 'crossing region' implies the
existence of a partner orbit and therefore a 
small number of non-relevant 'crossing regions'
has to be excluded.

To compute the contribution of the
described orbit pairs $({\bi x}_\gamma, {\bi x}_\gamma^p)$
to $K(\tau)$ we rearrange 
(\ref{K_tau}) into a sum over periodic
orbits and another sum over all the partners with small
action difference. This is based on the assumption
that the dominant off-diagonal contribution to  
$K(\tau)$, equation~(\ref{K_tau}), 
is due to the systematic correlation of actions
of the considered orbit pairs $({\bi x}_\gamma, {\bi x}_\gamma^p)$
while other correlations are negligible.
Then we sort the terms in the sums with respect to their
action differences.
Since the orbit length $T_\gamma$ is proportional to 
$T_H$ for fixed $\tau$,
one expects a large number of 'crossing regions'
for each orbit.
This allows us to replace the sum over action
differences $S_{\gamma,\bar{\gamma}}$
by an integral.
The first off-diagonal contribution
to $K(\tau)$ then reads
\begin{equation}
\label{K2_tau}
K^{(2)}(\tau) = 4 \tau \, \mbox{Re} \,
\left\langle \int\limits_0^\infty {\rm d}S \,
\left\langle \frac{{\rm d}N_{S,\gamma}}{{\rm d}S} 
\right\rangle_{(\gamma, T)}
\exp \left( i \frac{S}{\hbar} \right)  \right\rangle_{\Delta E}
\end{equation}
where $N_{S,\gamma}$ is the number of relevant 'crossing regions'
for a given periodic orbit ${\bi x}_\gamma$ with an associated action 
difference smaller than $S$.
In equation~(\ref{K2_tau}),
$\left\langle  \ldots \right\rangle_{(\gamma, T)}$ 
denotes a weighted average over all
orbits ${\bi x}_\gamma$ of given length
$T_\gamma = \tau T_H$. It is defined as
\begin{equation}
\label{avg_N}
\left\langle \frac{{\rm d}N_{S,\gamma}}{{\rm d}S} 
\right\rangle_{(\gamma, T)} \equiv
\frac{1}{T} \sum\limits_\gamma
\frac{{\rm d}N_{S,\gamma}}{{\rm d}S} |A_\gamma|^2
\delta (T-T_\gamma) \; .
\end{equation}
We proceed with the evaluation of (\ref{K2_tau}) by first
determining the geometry of the partner orbit ${\bi x}_\gamma^p$.
Then we show that the action
difference between ${\bi x}_\gamma$
and ${\bi x}_\gamma^p$ is indeed small if the orbit
${\bi x}_\gamma$ and its time-reversed version ${\bi x}_\gamma^i$
come close together in parts
of the phase space. Finally we derive the averaged number
$\left\langle {\rm d} N_{S,\gamma} / {\rm d}S 
 \right\rangle_{(\gamma,T_\gamma = T)}$
of relevant 'crossing regions' in the contributing regime
$S \lesssim \hbar$.

{\it Partner geometry.} To construct the partner orbit we analyse
the linearised equations of motion around
${\bi x}_\gamma$ in part ${\cal R}$ 
and around ${\bi x}^i_\gamma$ in section ${\cal L}$ in the
Poincar{\'e} surface of section (PSS), see figure
\ref{figure1}, and show
that a nontrivial solution representing the 
partner orbit ${\bi x}^p_\gamma$ exists under certain
conditions. The distance between the original orbit
${\bi x}_\gamma$ and its time-reversed
partner ${\bi x}^i_\gamma$ in the PSS
defined by the local transverse coordinates
at the phase space position 
${\bi x} \equiv {\bi x}_\gamma (t)$ is given by the 
vector
$\delta \vec{y} \equiv (\Delta x^\perp, \Delta p^\perp)$
\cite{Notation}.
In general, the time evolution of a small deviation
is determined by the starting point ${\bi x}_0$ of the
original path in phase space and the initial deviation
in the PSS $\delta \vec{y}_0$, e.g.
$\delta \vec{y} = \delta \vec{y} (t; {\bi x}_0, \delta \vec{y}_0)$.
Let us assume for the moment that the 'crossing region'
under consideration is characterized  by a small distance
$\delta \vec{y}$. Decomposing 
$\delta \vec{y}=y_u \vec{w}_u + y_s \vec{w}_s$
in terms of the
local unstable and stable manifolds $\vec{w}_{u,s}$ 
\cite{Gaspard98} with the expansion coefficients
\begin{equation}
\label{y_coefficients}
y_{u,s} \equiv \frac{\vec{w}_{s,u}^{\, T} \, Z \, \delta \vec{y}}
                 {\vec{w}_{s,u}^{\, T} \, Z \, \vec{w}_{u,s}} 
\quad \mbox{and} \quad
Z \equiv 
 \left( \begin{array}{cc} 0 & 1 \\ -1 & 0 \end{array} \right)
\end{equation}
the smallness of $\delta \vec{y}$ is given if
$|y_{u,s}| \ll 1$.
For simplicity we choose the relative orientation and lengths of
the $\vec{w}_{u,s}$ such 
that $(Z \vec{w}_s)^T \vec{w}_u = S_{{\rm spo}}$.
It will turn out that $|y_{u,s}| \ll 1$ is the relevant
regime for the evaluation
of $K(\tau)$ in the semiclassical limit.
The distance between  ${\bi x}_\gamma$
and the partner orbit ${\bi x}^p_\gamma$ at the
beginning of the first loop ${\cal R}$ is denoted by
$\delta \vec{x}_{R,i}$, see figure \ref{fig_pss}.
This vector lies in the PSS defined at the phase space position 
${\bi x} \equiv {\bi x}_\gamma (t)$
before the loop ${\cal R}$ [this corresponds
to the upper crossing region in figure \ref{fig_partner}].
Having passed loop
${\cal R}$ after time $T_R$ this distance has
changed to $R \; \delta \vec{x}_{R,i}$ with
$R$ being the stability matrix for loop ${\cal R}$.
Before (and after) the other part of the orbit
the difference between the time-reversed path
${\bi x}^i_\gamma$ and the partner ${\bi x}^p_\gamma$
is denoted by $\delta \vec{x}_{L^i,i}$ 
(and $L^i \; \delta \vec{x}_{L^i,i}$) where
$L^i \equiv F L^{-1} F$ is the stability matrix of
the time-reversed loop ${\cal L}$.
The matrix $F$ is defined as
\begin{equation}
F\equiv \left( 
  \begin{array}{cc} 1 & 0 \\ 0 & -1 \end{array} \right).
\end{equation}
Solving the linearised equations of motion under
the condition that the two parts of the partner orbit
fit together in the 'crossing regions'
yields the geometry of ${\bi x}_\gamma^p$
in terms of the distance $\delta \vec{y}$ 
between the original orbit ${\bi x}_\gamma$ and
its time-reversed ${\bi x}^i_\gamma$:
\begin{equation}
\label{geometry}
\delta \vec{x}_{R,i} = \left[ 1 - L^i R \right]^{-1}
\left[ 1 - L^i F \right] \delta \vec{y}.
\end{equation}
The corresponding condition for $\delta \vec{x}_{L^i,i}$
for the lower 'crossing region' in
figure \ref{fig_partner} is found in a similar way
\cite{Turek03a}.
This set of solutions $\delta \vec x$ includes
terms of order ${\cal O}(y_{u,s})$ and defines
the partner orbit for a given small 
$\delta \vec{y}$ representing a 'crossing region'.

We now argue that a 'crossing region' does not
yield a partner if the periodic orbit ${\bi x}_\gamma$
lies close to a self-retracing path during one of the
loops. This type of 'crossing region' is
described by a $\delta \vec{y}^{sr}$ such that
the original path ${\bi x}_\gamma$
and the time-reversed ${\bi x}^i_\gamma$
stay close together with $|y_{u,s}^{sr}| \ll 1$
holding during the entire loop ${\cal R}$.
The motion of the time-reversed path
${\bi x}^i_\gamma$ in ${\cal R}$ can 
then be obtained 
by linearisation around the original
${\bi x}_\gamma$ using the stability matrix $R$. In this
case one finds
$R \, \delta \vec{y}^{sr} = F \, \delta \vec{y}^{sr}$ neglecting
corrections of higher then first order in $y_{s,u}$.
The solution (\ref{geometry}) is then replaced by
$\delta \vec{x}_{R,i} = \delta \vec{y}^{sr}$ and
$\delta \vec{x}_{L^i,i} = 0$. It therefore
does not give a new partner orbit but just the
time-reversed periodic orbit ${\bi x}^p_\gamma = {\bi x}^i_\gamma$.
But contributions to $K(\tau)$ 
of this type are already treated
in the diagonal approximation and must not
be included in (\ref{K2_tau}).
However, because of the hyperbolic nature of
the dynamics the condition $|y^{sr}_{u,s}| \ll 1$
holds true for the entire loop ${\cal R}$ 
only if
the loop time $T^{sr}_R$ is smaller than a certain
minimal time $T_{R,{\rm min}}({\bi x}_\gamma(t), \delta \vec{y})$, 
i.e. $T_R^{sr} < T_{R,{\rm min}}$. Corresponding
arguments involving the time
$T_{L,{\rm min}}({\bi x}_\gamma(t), \delta \vec{y})$
hold for the other loop ${\cal L}$.
These minimal times $T_{(R,L),{\rm min}}$ are
determined by the time scale on which the linearization
breaks down and are implicitly given by the condition
\begin{equation}
\label{T_min}
 y_{u,s}(\pm T_{(R,L),{\rm min}}; {\bi x}, \delta \vec{y}) 
 = c_{u,s}({\bi x})
\end{equation}
where the $'+'$ ($'-'$) sign corresponds to
$y_u$ ($y_s$) and  $c_{u,s}({\bi x})$ are constants of 
order one.

Since (\ref{geometry}) uniquely defines the partner
orbit as sketched in figure \ref{fig_partner} one can show
that the Maslov index for the orbit ${\bi x}_\gamma$
equals that of ${\bi x}^p_\gamma$. The Maslov index of a periodic
orbit is given by the winding number of the stable or
unstable manifold \cite{Creagh90}. Since the partner orbit
is close to the original orbit in section ${\cal R}$
the contribution to the winding number accumulated
between the two 'crossing regions' is the same for both.
The second contribution
comes from loop ${\cal L}$ and is thus given by 
the geometry ${\bi x}^i_\gamma$. The total winding number 
for ${\bi x}^p$, the
sum of these two contributions, can be related to the
total winding number of ${\bi x}$ in the following way.
Time reversal symmetry implies the relation
$ {\vec w}_{s,u}({\bi x}_\gamma (t)) =
 - F {\vec w}_{u,s}({\bi x}^i_\gamma (T_\gamma-t))$
between the manifolds of ${\bi x}_\gamma$ and
${\bi x}^i_\gamma$. With this relation
one can show that the contributions
to the winding number coming from ${\bi x}_\gamma^i$ 
and  ${\bi x}_\gamma$ during ${\cal L}$ are equal.
Then the equality of the Maslov indices of
${\bi x}_\gamma$ and ${\bi x}^p_\gamma$ becomes
evident \cite{Turek03a} if one uses
similar arguments as in the
proof of the equality of the Maslov indices of
a periodic orbit and its time reversed 
counterpart \cite{Foxman97}. 

{\it Action difference.}
The geometry of the partner given by (\ref{geometry})
allows one to derive the action difference between the
two orbits of the pair $({\bi x}_\gamma, {\bi x}^p_\gamma)$
as a function of $\delta \vec{y}$.
Since the distance $\delta \vec{y}$ is assumed to be
small it is sufficient to expand the action
in $\delta \vec{y}$. However,
the expression
one obtains for $S$ \cite{Sieber02a} still contains
all the elements
of the stability matrices $R$ and $L$ because
the geometry of the partner orbit as given by (\ref{geometry})
depends on them. Since the existence of the
partner implies loop lengths larger than
$T_{(L,R),{\rm min}}$ the
vectors $\delta \vec{x}$ have to lie very
close to the respective local stable or unstable manifolds, e.g.
$\delta \vec{x}_{R,i} \approx y_s \vec{w}_s({\bi x})$.
This fact enables us to express
the action difference $S_{\gamma,\bar{\gamma}}$ between
the original orbit ${\bi x}_\gamma$ and the partner
orbit ${\bi x}^p_\gamma = {\bi x}_{\bar{\gamma}}$
in terms of the local manifolds \cite{Turek03a} and
the expansion coefficients $y_{u,s}$ given by
(\ref{y_coefficients}). Under the assumption that the
directions of the manifolds are continuous functions
of the position in phase space \cite{Gaspard98}
the result then reads
\begin{equation}
\label{action}
S_{\gamma,\bar{\gamma}} = S(\delta \vec{y}, {\bi x}) \approx
  (\vec{w}_u^T \, Z \, \vec{w}_s) \; y_s \, y_u
\end{equation}
which is correct up to second order in $\delta \vec{y}$.
In the semiclassical limit it is sufficient
to consider the regime given by
$|y_{u,s}| \sim \sqrt{\hbar / (\vec{w}_u^T \, Z \, \vec{w}_s)}
 = \sqrt{\hbar / S_{{\rm spo}}} \ll 1$
which justifies the above
restriction to small values of $|y_{u,s}|$.

The  equations (\ref{geometry}) representing the geometry of
${\bi x}^p$ and (\ref{action}) are invariant
under a shift of the PSS along the orbit within
a 'crossing region'.
This also implies that a 'crossing region' may include 
several selfcrossings in configuration
space and hence the number of partners is not
necessarily given by the number of selfcrossings as it was
the case for the uniformly hyperbolic
systems \cite{Sieber01,Sieber02}.


{\it Counting 'crossing regions'.}
Moving the PSS along ${\bi x}_\gamma$ 
each 'crossing region' is characterized by a
$\delta \vec{y}$ that starts close to ${\vec w}_s$
and ends almost parallel to ${\vec w}_u$.
The number of 'crossing regions' can thus
be determined
by counting how often the unstable components $y_u$
of the vectors $\delta \vec{y}$ go through a certain
fixed value $y_u^c$ as one moves along ${\bi x}_\gamma$.
This parameter $y_u^c$ fixes the
position within the 'crossing regions' used to identify
and count it. The total number of
'crossing regions' must not depend on $y_u^c$ and it will
be shown that this is indeed the case.


To evaluate the weighted average (\ref{avg_N})
of the number $N_S$ of contributing 'crossing regions' 
we make use of the following sum rule \cite{Parry90}
valid for ergodic systems
\begin{equation}
\label{sum_rule}
 \frac{1}{T} \sum\limits_\gamma |A_\gamma|^2
  \delta(T-T_\gamma) \int\limits_0^{T_\gamma} {\rm d}t
  f({\bi x}_\gamma (t)) \approx \int\limits_0^T {\rm d}t 
  f({\bi x}(t)) \quad \mbox{for} \quad T \to \infty.
\end{equation}
It relates the weighted average of
a function $f({\bi x})$ over all
periodic orbits of length $T$ to
a time average over a generic ergodic
trajectory ${\bi x}(t,{\bi x}_0)$ starting
at any ${\bi x}_0$ in phase space.
To apply (\ref{sum_rule}) in the calculation of
(\ref{avg_N}) one writes the number of events
where the time reversed path comes close to the
original one as
\begin{equation}
\label{local_number}
 \frac{{\rm d}^2 N_{S,\gamma}({\bi x}_\gamma(t))}{{\rm d}S \; {\rm d}t} = 
  \rho({\bi x}_\gamma(t), S, y_u^c)
  \frac{\dot y_u(0; {\bi x}_\gamma(t), S, y_u^c)}
       {y_u(0; {\bi x}_\gamma(t), S, y_u^c)},
\end{equation}
where $\rho({\bi x}_\gamma(t); S, y_u^c)$ is the density
of partners per action in the PSS located at
${\bi x}_\gamma(t)$. The ratio $\dot{y}_u / y_u$ describes
the velocity of the flow in the PSS so that
(\ref{local_number}) indeed gives the number
of partners per action and time. In expression
(\ref{local_number}) the position in the PSS 
is specified in terms of $S$ and $y_u^c$ 
using (\ref{action}).

The density $\rho({\bi x}; S, y_u^c)$ is determined
by the probability that the time reversed path
goes through the point $(S,y_u^c)$ in the
PSS defined at ${\bi x}$.
The long time limit is thus
given by the ergodic density $\rho_0 = T/\Sigma(E)
=\tau / (2 \pi \hbar)$ with $\Sigma(E) = (2 \pi \hbar)^2 \bar{d}(E)$
being the volume of the energy surface in phase space.
However, since certain 'crossing regions' with
short loop lengths
characterized by $T_{(R,L)} < T_{(R,L),{\rm min}}$,
see (\ref{T_min}),
do not yield a partner one has to exclude
parts of the time reversed path of length
\begin{equation}
\label{T_min_total}
T_{{\rm min}}({\bi x}, S) \equiv
 2[T_{R,{\rm min}}({\bi x}, S, y_u^c) + T_{L,{\rm min}}({\bi x}, S, y_u^c)]
\end{equation}
with the factor $2$ coming from time reversal symmetry.
According to definition (\ref{T_min}) this
time $T_{{\rm min}}({\bi x},S)$ is determined by the time
it takes for the unstable component $y_u$ to grow
from the small value $S/(S_{\rm spo} c_s) \ll 1$
to the value $c_u \sim 1$. It therefore
does not depend on $y_u^c$.
To compute 
(\ref{avg_N}) we first apply (\ref{sum_rule}) to
(\ref{local_number}).
Then the minimal time (\ref{T_min_total}) is
of the same order as the Ehrenfest time and
in the regime of 
small action differences $S \sim \hbar$
given by the asymptotic expression
\begin{equation}
\lambda T_{{\rm min}}({\bi x}, S) \approx 2 \ln \left[
  c_u({\bi x}) c_s({\bi x}) \frac{S_{{\rm spo}}}{S}
  \right] \gg 1
\end{equation}
where we used the standard definition of the
Lyapunov exponent $\lambda$ in terms of
any long ergodic path \cite{Gaspard98}.
Using this relation we find that the main contribution
to (\ref{K2_tau}) in the semiclassical limit is given
by the averaged number of 'crossing regions'
\begin{equation}
\label{result_N}
\left\langle \frac{{\rm d} N_{S,\gamma}}{{\rm d}S} 
 \right\rangle_\gamma = \rho_0
 \left[ \lambda T + 2 \ln \frac{S}{S_{{\rm spo}}} - {\rm const}
   \right] \; ,
\end{equation}
where ${\rm const}$ depends on the structure of
the phase space of the considered system but is independent
of the action difference $S$ and the time $T$ \cite{Turek03a}.
The first of the three contributions on the r.h.s.\ of
(\ref{result_N}) is the largest one ($\sim \hbar^{-2}$) 
and represents the ergodic properties of the system. The
second term $\sim \hbar^{-1} \ln \hbar$
reflects the underlying dynamics of
the system \cite{Turek03} and
is much smaller than the first but still
logarithmically larger than the third term $\sim \hbar^{-1}$.
Inserting (\ref{result_N}) into (\ref{K2_tau})
one finds that it is this logarithmic correction to the
ergodic behaviour that gives the RMT result
\begin{equation}
K^{(2)}(\tau) \approx -2 \tau^2.
\end{equation}


{\em GOE-GUE transition.} The crossover between the 
universality classes as time reversal symmetry is
broken has been originally semiclassically obtained in Ref.\
\cite{Bohigas95} within the diagonal approximation. Here we
summarize a dynamical evaluation of this transition for the
first off-diagonal correction. The appropriate transition
parameter $\alpha$ interpolating between GOE and GUE is
given by the ratio between the root mean square of a typical
time-reversal symmetry breaking matrix element and the mean
level spacing \cite{Bohigas95}. We shall consider the case
where the symmetry is broken by a uniform magnetic field $B$
perpendicular to a uniformly hyperbolic 
two-dimensional system. Since for
an orbit pair  $({\bi x}_\gamma, {\bi x}^p_\gamma)$
one of the common loops (${\cal L}$ or ${\cal R}$) between the 
'crossing regions' are traversed in opposite direction in configuration
space, the orbit pair acquires, owing to the overall magnetic 
flux enclosed, an additional action difference
$4\pi {\cal A} B / \phi_0$. Here $ {\cal A}$ is the 
enclosed (directed) area of the loop and $ \phi_0 = h/2e$ the 
flux quantum. For hyperbolic systems the distribution of 
directed areas enclosed by trajectories of length $t$ is,
to good approximation\cite{Richter00}, 
Gaussian with variance $t\beta$ where $\beta$ is a system
specific parameter. To compute $K^{(2)}(\tau)$ at finite
$B$ we must additionally integrate, for given loop length
$t$, the flux-induced action differences over the
Gaussian area distribution. This results in a further damping  
$\exp[-t/t_B]$ with
$ 1/t_B = 2\beta (2\pi B/\phi_0)^2$.
Counting the 'crossing regions' with this
additional weight eventually gives, together
with the diagonal term, \cite{Turek03a,Nagao02}
\begin{equation}
K(\tau;\alpha) \approx  \tau [1 + (1-2\tau)\e^{-8\pi^2 \alpha^2 \tau}]
\qquad \mbox{for} \qquad \tau \to 0
\end{equation}
with $\alpha^2 \tau = (B/\phi_0)^2 \beta T$.
This precisely coincides with the form factor of parametric RMT 
\cite{Pandey83} in the short time limit.

To conclude, we have shown how correlations in the
action of classical paths determine the spectral statistics
of the quantum mechanical energy eigenvalues. We
derived the next to leading order contribution for the
spectral form factor $K(\tau)$ and showed that it is
identical to the corresponding RMT result. 
Our derivation is canonical invariant and based on phase 
space arguments only.  This leads to the result that
correlations in the classical action
in hyperbolic chaotic systems are
caused by 'crossing regions' in phase space
where an orbit and its time-reversed
version come close together. Since
our method avoids the concept of crossings in
configuration space it is suited to be extended
to systems with more than two degrees of freedom.


\ack
We acknowledge support from the DFG under contract Ri 681/5-1. 
We thank particularly M. Sieber for many helpful comments and
M. Brack, S.~M{\"u}ller,
P. Schlagheck, and J.D. Urbina for enlightening
discussions. 

\vspace{0.6cm}
{\em Note added.} -- After we finished this paper we
received a preprint by D.Spehner which contains some
similar material.





\begin{thebibliography}{99}
\bibitem{Notation}
  For clarity of notation we
  write vectors in the PSS using an arrow 
  (e.g. $\delta \vec{y}$) in contrast to
  vectors in phase space (e.g. ${\bi x}$).
\bibitem{Turek03} 
  The crucial logarithmic dependence has been numerically confirmed
  for the cardioid billiard \cite{Turek03a}
\end{thebibliography}
\end{document}